\documentclass{jltp} 
\usepackage{graphicx}
\usepackage{calc}
\usepackage{amsmath}
\usepackage{amssymb}
\usepackage{bm}

\bmdefine\bomega{\omega}
\bmdefine\bOmega{\Omega}
\bmdefine\bnabla{\nabla}
\bmdefine\bkappa{\kappa}

\begin{document}

\def\topfraction{1} \def\textfraction{0}

\title{\boldmath Measurement of Turbulence in Superfluid $^3$He-B} 

\author{A.P.~Finne$^{*}$, S.~Boldarev$^{*,\dagger}$, V.B.~Eltsov$^{*,\dagger}$, and
M.~Krusius$^*$}

\address{$^*$Low Temperature Laboratory, Helsinki University of
  Technology,\\ P.O.
  Box 2200, FIN-02015 HUT, Finland\\
  $^\dagger$Kapitza Institute for Physical Problems, 119334 Moscow, Russia}
\vspace{-6mm}

\runninghead{A.P.~Finne \textit{et al.}}{Measurement of Turbulence in
  Superfluid $^3$He-B}

\maketitle

\vspace{-6mm}

\begin{abstract}
The experimental investigation of superfluid turbulence in $^3$He-B is generally not possible with the techniques which have been developed for $^4$He-II. We describe a new method by which a transient burst of turbulent vortex expansion can be generated in $^3$He-B. It is based on the injection of a few vortex loops into rotating vortex-free flow. The time-dependent evolution of the quantized vorticity is then monitored with NMR spectroscopy. Using these techniques the transition between regular ({\it i.e.} vortex number conserving) and turbulent vortex dynamics can be recorded at $T \sim 0.6\,T_{\rm c}$ and a number of other characteristics of turbulence can be followed down to a temperature of $T \lesssim 0.4\,T_{\rm c}$.

\end{abstract} \vspace{-12mm}

\section{INTRODUCTION}
\label{Introduction}

Quantum turbulence, the seemingly chaotic motion of quantized vortex lines in a disordered network, is a prominent characteristic of superfluid $^4$He-II, where it has been known to exist for almost half a century.\cite{VinenNiemela} Up to recently, $^4$He-II was the only superfluid system in which this phenomenon has been investigated. $^4$He-II is believed to display turbulence in the entire superfluid temperature range which so far has been probed with measurements. In superconductors friction in vortex motion is always high and turbulence is not observed: Only in the extreme clean limit one might hope to see any signs of other behavior.  At present the closest superconducting analogues are dynamically driven vortex avalanches.\cite{Johansen}

Superfluid $^3$He is an intermediate case between these two extremes. The dynamics of vortex motion in the $^3$He superfluids has primarily been studied in rotating cryostats. In the past the generally accepted view was that the high mutual friction leads to exponentially damped motion and forces vortices to evolve rapidly along well defined trajectories, so that the number of vortex lines remains conserved in dynamic processes. A large number of measurements over the years has proven this to be the case in both $^3$He-A and $^3$He-B at $T>0.6\,T_{\rm c}$.\cite{CollectiveMotion}

Even at low temperatures ($\sim 0.2\,T_{\rm c}$) a duplication of the
classic Vinen-vibrating-wire measurement\cite{VortexPrecession} verified
that in $^3$He-B stable vortex motion persists for hours in the absence of
applied flow. In this experiment superfluid circulation was trapped around a
thin wire suspended along the symmetry axis of a cylindrical container
while the cryostat was rotating. When rotation was stopped, the trapped
circulation started to unwind while a vortex filament, stretched between
the wire and the cylinder wall, precessed around the wire and spiraled down
along the whole length of the 15\,mm long cylinder. The precessing spiral
motion lasted for more than 35\,h and the precession frequency remained
constant with a precision of 0.5\%.

The first indications of rapid non-linear vortex proliferation in $^3$He-B came from measurements of the critical velocity of vortex formation at temperatures $\lesssim 0.6\,T_{\rm c}$ (Ref.~\onlinecite{Ruutu}, see p. 140--141). Instead of the usual single-vortex formation at a reproducible critical velocity, sudden avalanches of vortices were observed in accelerating rotation. These bursts were interpreted to mean that at lower temperatures mutual-friction damping had decreased sufficiently so that turbulence became possible.

The first evidence for tangled vorticity was reported from vibrating wire measurements in a non-rotating cryostat at temperatures below $0.2\,T_{\rm c}$.\cite{AndreevReflection} A vortex network was found to be produced with a vibrating wire resonator when it was driven at high level above a critical velocity of vortex formation. This was inferred by studying the damping of a second wire, vibrating at low level and probing the density of quasiparticle excitations which are Andreev retro-reflected by the superflow fields in the vortex tangle. Similar later measurements have allowed a determination of the vortex density in the wire-generated network. At zero externally applied pressure and at a temperature of $0.12\,T_{\rm c}$  the tangle localized around the vibrating wire turned out to have extremely low density, corresponding to an average inter-vortex distance of 0.2\,mm, and to decay away rapidly in about 5\,s, when the generator wire was switched off.\cite{LineDensity}

There exist no immediately obvious techniques by which turbulence can be easily generated and detected in $^3$He-B. The above two examples suggest two ways which are supplementary, since they work in different temperature regimes. As outlined below in Sec.~\ref{Experiment}, NMR on a rotating sample can be used at temperatures down to about $0.4\,T_{\rm c}$. In the zero temperature limit, where the density $\rho_{\rm n}$ of the normal component approaches zero, moving objects can be used to generate vortices and, when driven in a high-quality resonance mode, to detect them. In this report we outline the techniques by which turbulence is generated and detected in the intermediate temperature regime 0.4 -- $0.6\,T_{\rm c}$, where mutual friction damping and $\rho_{\rm n}$ are still finite. Recently a practical technique was discovered for injecting vortex loops into vortex-free flow.\cite{KH} It then became possible to monitor the nature of vortex dynamics as a function of temperature and flow velocity in a controlled fashion.\cite{Turbulence} The transition, which separates regular vortex motion from turbulence, could now be mapped. The foundation, on which the interpretation of these results is built, was laid down by an earlier set of important hydrodynamic measurements\cite{Bevan} on mutual friction which provided both the dissipative and reactive mutual friction coefficients $\alpha(T,P)$ and $\alpha^{\prime}(T,P)$ as a function of temperature $(T)$ and pressure $(P)$. It turns out that in $^3$He-B the variation of mutual friction with temperature happens to be such that the transition falls in the middle of the experimental temperature range, to 0.5 -- $0.6\,T_{\rm c}$.

These techniques have been employed in three recent studies investigating turbulence in $^3$He-B.\cite{FlightTime,NeutronTurbulence,TurbPhaseDiagram} The first\cite{FlightTime} reports on the flight time of the vorticity when it expands along the rotating column.
The second\cite{NeutronTurbulence} describes how turbulence starts in
rotating vortex-free flow in the presence of neutron radiation. The third\cite{TurbPhaseDiagram} discusses the mutual-friction dependence of the transition from
regular to turbulent vortex dynamics, {\it i.e.} the transition line as a function of $T$ and $P$ at high flow velocities, where the transition is velocity independent. These transition properties will also be illuminated in Sec.~\ref{TurbTransition} of this report, but from a different point of view. NMR measurements from the transient period, when the turbulent vorticity decays into the equilibrium rotating state of rectilinear vortex lines, will be analyzed in a forthcoming report.\cite{ToBePublished}

According to current view no similar kind of turbulence as in $^4$He-II or $^3$He-B exists in the anisotropic superfluid $^3$He-A. Here vortex motion is highly damped at all experimentally accessible temperatures and fast vortex motion is supported by the appearance of a different structure of vorticity, the vortex sheet.\cite{VortexSheet}

\section{ROLE OF DAMPING IN SUPERFLUID TURBULENCE} \label{Analysis}

\begin{figure}[t]
\centerline{\includegraphics[width=0.9\linewidth]{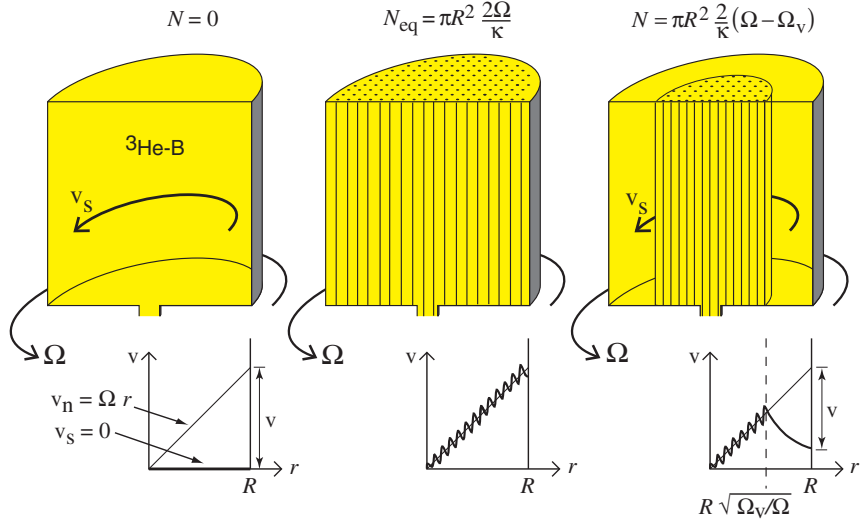}}
\caption{Different states of uniformly rotating superfluid.
  {\it (Left)} The vortex-free state is the highest energy state and the
  initial state in these measurements, with the maximum applied counterflow
  velocity at given rotation $\Omega$: $v = v_{\rm n} - v_{\rm s} = \Omega
  r$. {\it (Middle)} The equilibrium vortex state is the lowest energy
  state. Here an array of rectilinear vortex lines fills essentially all of
  the container and the superfluid component is on an average in solid body
  rotation, {\it i.e.} stationary in the rotating frame. The number of
  vortex lines is $N_{\rm eq} \approx \pi R^2 \, 2\Omega/\kappa\sim 10^3$
  if $\Omega \sim 1\,$rad/s. This is the final state after injection of
  seed vortex loops, if the vortex expansion has been turbulent. {\it
    (Right)} A meta-stable vortex cluster may include any number of
  rectilinear vortex lines, $0 < N < N_{\rm eq}$. Within the central
  coaxial vortex cluster the areal density of the lines $2\Omega/\kappa$
  corresponds to solid body rotation. The counterflow velocity outside the
  cluster is $v = \kappa (N_{\rm eq} - N)/(2\pi r)$, where we define $N = \pi R^2 \, 2\Omega_{\rm v}/\kappa$. This is the final
  state after injection, if loop expansion has been regular and the number
  of individual vortices has been conserved. In this case $N \lesssim 30$
  after a Kelvin-Helmholtz injection event (Fig.~\ref{StepDistribution}).}
\label{RotSupFluidStates}
\end{figure}

Let us consider a hydrodynamic equation which describes the evolution of quantized vorticity in a limit, which becomes useful in the interpretation of our measurements. We apply dimensional analysis to this equation in the same manner as one characterizes different flow regimes with the Navier-Stokes equation in viscous hydrodynamics. This immediately leads to the notion that a change in the nature of vortex motion is expected as a function of mutual friction damping. The argument is simple-minded, but makes the suppression of turbulence by mutual friction damping understandable in superfluid hydrodynamics.

The equation of motion for a viscous incompressible fluid is the
Navier-Stokes equation,
\begin{equation}\label{navierstokes}
    \frac{\partial \mathbf{v}}{\partial t}+(\mathbf{v}\cdot
    \bnabla)\mathbf{v} = \mathbf{F}/\rho-\bnabla P/\rho+
    \nu\Delta\mathbf{v}\,,
\end{equation}
where $\mathbf{F}$ is an external force per unit volume, $P$ pressure, and
$\nu$ kinematic viscosity. In order to compare flows with different
viscosities it is useful to write the Navier-Stokes equation in
dimensionless units $x^0=x/L$, $v^0=v/V$ and $t^0=Vt/L$, where $L$ and $V$
are the characteristic length and velocity scales. The relative importance
of the inertial term $(\mathbf{v}\cdot\bnabla)\mathbf{v} \sim V^2/L$ and
the viscous term $\nu\Delta\mathbf{v}\sim\nu V/L^2$ is used to characterize
the flow. The ratio of these two terms is called the Reynolds number
$Re=(V^2/L)/(\nu V/L^2)=VL/\nu$. For flows with small $Re$ the dissipative
forces are dominant and the resulting flow is typically laminar. If $Re$ is
large, the inertial forces are dominant and the flow is usually turbulent.

The same type of classification in superfluid hydrodynamics can be based on an equation
for the ``coarse-grained'' superfluid velocity $\mathbf{v}_{\rm s}$. It is obtained by
averaging over volumes containing many vortex lines. The
equation\cite{Sonin} is constructed from the Euler equation:
\begin{equation}\label{regular_euler_force}
\frac{\partial \mathbf{v}_{\rm s}}{\partial t}
+(\mathbf{v}_{\rm s}\cdot \bnabla)\mathbf{v}_{\rm s}
=\mathbf{F}/\rho_{\rm s}-\bnabla \mu,
\end{equation}
which can also be written as
\begin{equation}
\frac{\partial \mathbf{v}_{\rm s}}{\partial t} +
\bomega \times \mathbf{v}_{\rm s} =
\mathbf{F}/\rho_{\rm s}-\bnabla\biggl(\mu + \frac{v_{\rm s}^2}{2}\biggr),
\label{eq:euler2}
\end{equation}
where $\mu$ plays the role of the chemical potential and $\bomega = \bnabla
\times \mathbf{v}_{\rm s}$ is the vorticity.

For such coarse-graining to be meaningful the volume, over which we average, should
contain many roughly similarly oriented vortex lines. As each vortex carries a fixed quantized
circulation $\kappa = 6.61 \cdot 10^{-4}\,\mbox{cm}^2/\mbox{s}$ the
number of vortex lines in the sample can be estimated from below as $[\langle
\omega \rangle/\kappa]  L^2 \sim V L / \kappa$. We call this
combination the ``superfluid Reynolds number'' $Re_{\rm s} = V L / \kappa$ and
require that  $Re_{\rm s} \gg 1$ for the coarse-grained equation
\eqref{regular_euler_force} to be applicable.

The Euler equation basically expresses the conservation of energy and Eq.~\eqref{eq:euler2} thus accounts for the kinetic energy of the global
flow. The vortex line tension, which is associated with the $1/r$ local
velocity profile around the vortex cores, is neglected. The tension can be
included by introducing a correction\cite{Sonin,Glaberson} to the superfluid
velocity $\delta \mathbf{v}_{\rm s} = (\kappa/4\pi) \ln (b/a) \bnabla
\times \hat\bomega$, where $b$ is roughly the intervortex spacing, $a$ is the
vortex core size, and $\hat\bomega$ is the unit vector in the direction of
$\bomega$. Dimensional arguments give the estimate $v_{\rm s} /
\delta v_{\rm s} \sim V/(\kappa/L) = Re_{\rm s}$. Thus for values $Re_{\rm
  s}\gg 1$, the tension can be neglected.

In the classical equation \eqref{navierstokes} the term $\mathbf{F}$ represents
some force applied externally to the fluid. In the case
of superfluids there exists an intrinsic contribution to $\mathbf{F}$, the mutual
friction force $\mathbf{F}_{\rm mf}$, which arises from the interaction of
vortex lines with normal excitations. This force, acting on a single
vortex line element, is\cite{Sonin}
\begin{equation}\label{mfriction}
    \mathbf{F}_{\rm mf}'=-\alpha\rho_s\hat\bkappa\times
     [\bkappa\times(\mathbf{v}_{\rm n}-\mathbf{v}_{\rm sl})]-
     \alpha'\rho_s\bkappa \times(\mathbf{v}_{\rm n}-\mathbf{v}_{\rm sl}),
\end{equation}
where $\alpha(T,P)$ and $\alpha'(T,P)$ are the dissipative and reactive
mutual friction coefficients, $\mathbf{v}_{\rm sl}$ is the local superfluid velocity at the position of the vortex line, $\bkappa$ is directed along the vortex line and
$\hat\bkappa$ is the corresponding unit vector. In general the averaging of the non-linear force in Eq.~\eqref{mfriction} over volumes containing many vortex lines cannot be done without
knowledge of the vortex configuration. Here for simplicity we assume a
locally polarized vortex tangle,\cite{volovik2reg} where all vortex lines
in the averaged volume have the same direction $\bkappa$. In this case the
transformation from $\mathbf{F}_{\rm mf}'$ to the coarse-grained force in
equation \eqref{eq:euler2} is done by replacing $\bkappa$ with $\bomega$
and $\mathbf{v}_{\rm sl}$ with $\mathbf{v}_{\rm s}$. Eq.~\eqref{eq:euler2} thus becomes
\begin{equation}
\frac{\partial \mathbf{v}_{\rm s}}{\partial t} +
\bnabla\biggl(\mu + \frac{v_{\rm s}^2}{2}\biggr) =
\mathbf{v}_{\rm s} \times \bomega
+ \alpha' \bomega \times (\mathbf{v}_{\rm s} - \mathbf{v}_{\rm n})
+ \alpha \hat\bomega \times [ \bomega \times
(\mathbf{v}_{\rm s} - \mathbf{v}_{\rm n})]\,.
\label{eq:dyn}
\end{equation}
This equation has traditionally been used to examine small deviations of
vortex lines from equilibrium configurations. We use it to
highlight the role of mutual friction in superfluid
turbulence.

The viscosity of the normal component in superfluid $^3$He is large: $\nu
\sim 1\, \mbox{cm}^2/\mbox{s}$. As a result, the normal component is in practice always
in well-defined externally imposed laminar motion. The velocity, which is induced
in the normal component by the mutual friction force, can be ignored in flows with
large $Re_{\rm s}$: From Eq.~\eqref{navierstokes}, by
substituting $\mathbf{F}$ with the mutual friction force, we find
that $v_{\rm n}/v_{\rm s} \sim (\kappa/\nu) Re_{\rm s}$ (in the
intermediate temperature range where $\rho_{\rm n} \sim \rho_{\rm s}$ and
$\alpha \sim 1$). Thus the influence of the mutual friction force on the
normal component can be ignored up to relatively fast motion of the
superfluid component with $Re_{\rm s} \sim 2\cdot 10^3$. Even beyond that
velocity (which has not been experimentally reached yet) the motion of the
normal component remains laminar as $Re_{\rm n} \sim (\kappa/\nu)^2 Re_{\rm
  s}$. (Note that these conclusions do not apply to superfluid $^4$He where
$\nu \sim \kappa$.)

In rotating experiments $\mathbf{v}_{\rm n} = \bOmega \times \mathbf{r}$ and we call $\mathbf{v} = \mathbf{v}_{\rm s}- \mathbf{v}_{\rm n}$ the {\it counterflow velocity} (Fig.~\ref{RotSupFluidStates}). To simplify Eq.~\eqref{eq:dyn} we change to the rotating reference frame (where $v_{\rm n} = 0$ and which we assume to be inertial):
\begin{equation}\label{super_euler_final}
\frac{\partial \mathbf{v}_{\rm s}}{\partial t} +
\bnabla\biggl(\mu + \frac{v_{\rm s}^2}{2}\biggr) =
(1-\alpha')(\mathbf{v}_{\rm s} \times \bomega)
+ \alpha \hat\bomega \times [ \bomega \times
\mathbf{v}_{\rm s}]\,.
\end{equation}
By applying dimensional analysis to this equation, we can construct a
quantity, which has similar physical meaning as the Reynolds number in the case of the
Navier-Stokes equation: The ratio of the inertial
$(1-\alpha')(\mathbf{v}_{\rm s} \times \bomega) \sim (1-\alpha')V^2/L^2$
and dissipative $\alpha \hat\bomega \times [ \bomega \times \mathbf{v}_{\rm
  s}] \sim \alpha V^2/L^2$ terms is here $1/q=(1-\alpha')/\alpha$. Similar
to the role of the Reynolds number in classical hydrodynamics we might
expect that when $1/q$ is large the flow is turbulent and when $1/q$ is
small the flow is laminar.\cite{Turbulence} Unlike in the classical case,
here only the intrinsic properties of the fluid, $\alpha(T,P)$ and $\alpha'(T,P)$,
determine the nature of the flow, since $1/q$ does not depend on the externally imposed geometry or the velocity of the flow.

\section{PRINCIPLE OF MEASUREMENT} \label{Experiment}

None of the established methods to generate turbulence in $^4$He-II are applicable in $^3$He-B in the intermediate temperature regime. A new approach is therefore essential. Fortunately compared to $^4$He-II, other possibilities are available. One major difference is that vortex formation can be experimentally controlled in $^3$He-B to much better degree. If remanent vortices and other extrinsic sources of vortex formation can be eliminated, then a high energy barrier prevents vortex formation via intrinsic mechanisms up to relatively high counterflow velocities, typically of order 1\,cm/s. Technically the simplest means of generating high vortex-free flow velocities is to rotate the sample container. Vortex loops can then be injected in the vortex-free flow by means of a few different techniques and their dynamic evolution can be monitored with NMR. Fig.~\ref{RotSupFluidStates} classifies schematically the different rotating states, from the initial vortex-free state to the equilibrium vortex state.

The most versatile injection method at present time is the Kelvin-Helmholtz instability of the AB phase boundary.\cite{KH} Here the injection can be performed at variable values of rotation $\Omega$ and temperature $T$. This requires a two-phase sample, like that shown in Fig.~\ref{ExpSetUp} on the left, where the center section of the long sample cylinder is maintained with a magnetic barrier field in the A phase.

To study the transition from regular to turbulent vortex dynamics as a function of temperature and flow velocity, we need a three-step process: 1) to set up the vortex-free rotating state, 2) to trigger the Kelvin-Helmholtz instability, and 3) to distinguish the type of final state. For controlling the experiment with the NMR measurement it needs to be able 1) to verify that we are  correctly in the vortex-free initial state, 2) to indicate that the injection is properly executed, and 3) to determine the number of rectilinear vortex lines in the final state. Surprisingly it turns out that the transition to turbulence is so sharp that the final state contains either a few rectilinear vortex lines (Fig.~\ref{RotSupFluidStates} right), if the temperature is above the transition, or else the equilibrium number of vortex lines (Fig.~\ref{RotSupFluidStates} middle). As seen from the NMR absorption spectra in Fig.~\ref{ExpSetUp} on the right, the line shapes of these two final states differ so profoundly that the classification in regular and turbulent events can be performed by inspection only. This is the case if the rotation velocity $\Omega \gtrsim 0.5\,$rad/s and the so-called  ``counterflow absorption peak'' (CF peak), generated by the vortex-free flow, is well developed.

In this report we use the KH instability as injection method. It is triggered with a small step increase in rotation $\Delta \Omega$ which is typically a few percent of the total rotation. This means that in our measurement the externally controlled conditions are maintained constant, since practically even the rotation velocity can be regarded as constant throughout the measurement. This amounts to an important simplification of the analysis. The time-dependent evolution of the injected vorticity lasts from tens of seconds to minutes, depending on temperature and rotation velocity.  During this transient period NMR signals can be measured which yield information on the dynamics. One such measurement is the determination of the flight time $\tau_{\rm F}$ of the vorticity from the AB interface to a  detector coil,\cite{FlightTime} as illustrated for the lower half of the sample container in Fig.~\ref{ExpSetUp}. Another measurement involves the configuration in which the vorticity propagates along the rotating column.\cite{ToBePublished}

The injection of vortex loops can be accomplished also by other means. In Ref.~\onlinecite{NeutronTurbulence} neutron irradiation was discussed for injection. In Refs.~\onlinecite{Ruutu} and \onlinecite{VorFlow} a sudden burst of vortex loops through the orifice on the bottom of the sample space was observed to start the turbulence.
\vspace{-5mm}

\subsection{Sample Setup}\label{Sample}

\begin{figure}[tp]
\centerline{\includegraphics[width=1.0\linewidth]{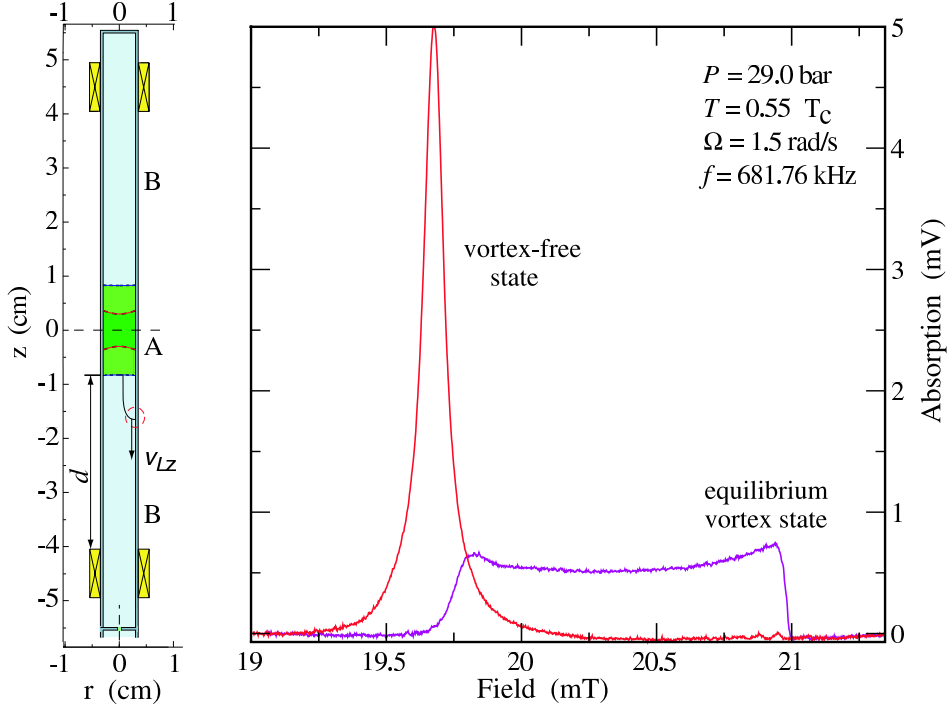}}
\caption{$^3$He sample and NMR measurement. {\it (Left)} The sample
  container is a quartz tube with diameter 6\,mm and length 110\,mm. It is
  separated from the rest of the liquid $^3$He volume with a partition disc
  at the bottom of the sample space.  In the disc an orifice of 0.75\,mm
  diameter provides the thermal contact to a liquid column which connects
  to the sintered heat exchanger on the nuclear cooling stage. Three
  superconducting coil systems produce independently axially oriented
  magnetic fields along different sections of the tube: at both ends are
  the polarization fields for NMR detection and in the middle the barrier
  field for stabilizing $^3$He-A. The dashed curves, approximately
  symmetric about the middle of the sample tube, show the locations of
  the AB interfaces in the gradient of the barrier solenoid when its
  current is $I_{\rm b} = 4\,$A (short section of A phase with parabolic
  contour of AB interfaces) and $I_{\rm b} = 8\,$A (longer section of A
  phase with almost flat interfaces).\protect\cite{FootNote} {\it (Right)} Two
  NMR absorption spectra in rotation. In the equilibrium vortex state the
  line shape borders steeply on the right to the Larmor field at 21.0\,mT,
  while at low values of field sweep it extends to the temperature
  dependent cut off of the flare-out texture, here at 19.7\,mT.  In the
  vortex-free state the line shape displays a large ``counterflow peak''.
  Its maximum is located at the cut-off value of the ``flare-out texture''.
  Both line shapes have been measured at the same temperature and have
  equal integrated absorptions. } \label{ExpSetUp}
\end{figure}

The liquid $^3$He sample and the NMR measurement have been summarized in Fig.~\ref{ExpSetUp}. Additional details have been explained in Refs.~\onlinecite{Rob} and \onlinecite{NeutronTurbulence}. The long sample tube is surrounded by three superconducting end-compensated coil systems. These supply the two constant field regions at both ends of the tube for NMR polarization and the barrier field to maintain the A phase in the center. If the barrier field $H_{\rm b}$ exceeds the thermodynamic B$\rightarrow$A transition field $H_{\rm AB}$, two AB interfaces are formed symmetrically around the center of the sample tube. Their shapes and locations depend on the current $I_{\rm b}$ in the barrier magnet and on the externally controlled variables $T$, $P$, and $\Omega$.\cite{AB-InterfaceShape} As an example, two locations of the AB interfaces are shown in Fig.~\ref{ExpSetUp}. For the measurement of the turbulent transition the locations of the AB interfaces are not of importance. However, to determine the flight time $\tau_{\rm F}$ the distance $d$ has to be accurately known. It is obtained from the calculated field profile of the solenoidal magnet system.

The continuous wave NMR measurement is performed at constant frequency $f$, using a linear sweep of the axially oriented polarization field $H$. The two absorption spectra in Fig.~\ref{ExpSetUp} display the extremes  of different possible line shapes. In fact, to monitor the transition to turbulence, they turn out to be the only two line shapes needed. The large peak in the vortex-free state is produced by the orienting effect of the counterflow on the order parameter texture and is called the {\it counterflow peak}. Its shift from the Larmor field (at 21.02\,mT) is used for temperature measurement and its height is a sensitive function of the number of vortex lines.

When a cluster of rectilinear vortex lines is formed, the CF peak height is reduced. The reduction is linear with vortex number $N$ in the limit $N \ll N_{\rm eq}$, if all other variables remain constant. At large values of $N$ the reduction becomes nonlinear and finally, if $N$ is continuously increased until it reaches the maximum possible value $\approx N_{\rm eq}$, the spectrum looks entirely different: Here much of the absorption is shifted to high fields and borders prominently to the Larmor edge at 21.0\,mT. With increasing $N$ the intensity in the CF peak is thus shifted closer to the Larmor edge. Therefore one may monitor the increase in $N$ by recording either the CF peak height or the absorption at the Larmor edge. In the latter region the absorption is less affected by temperature changes and often it is more practical to monitor the evolution of the vortex expansion process at fixed polarization field value close to the Larmor edge.

\begin{figure}[t]
\centerline{\includegraphics[width=0.9\linewidth]{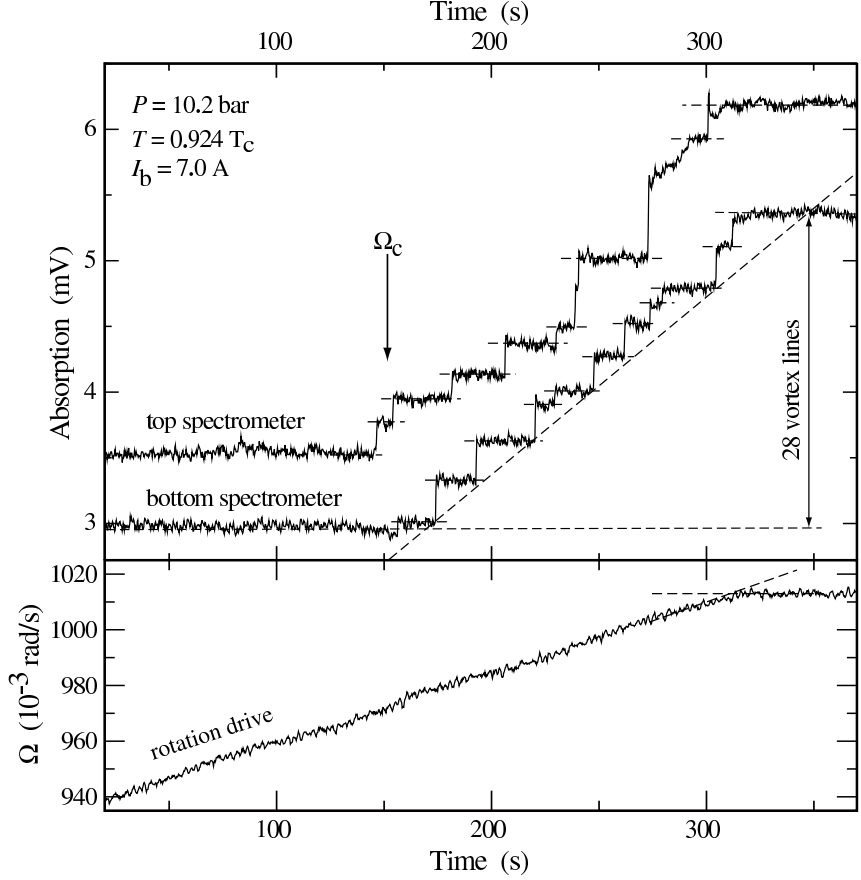}}
\medskip
\caption{Absorption response during slow linear increase of rotation while a sequence of the $\sim 10$ first Kelvin-Helmholtz instability events is traversed. {\it (Top)} The
  measurement is performed at high temperatures where vortex motion is regular. Here
  the number of vortex loops injected in the B phase equals the number of
  new rectilinear vortex lines which are added to the B phase sections per
  instability event. The measurement monitors the increase in rectilinear
  vortex lines per event. The increase is proportional to the height of the
  discontinuous steps in the linear staircase patterns and, as seen from
  the step heights, varies randomly per event. The NMR absorption is
  measured at constant polarization field at a value just below the Larmor
  edge ({\it e.g.} for the bottom spectrometer $H = 20.97$\,mT , see
  Fig.~\protect\ref{ExpSetUp}).  {\it (Bottom)} The rate of
  rotational acceleration is $2 \cdot 10^{-4}\,$rad/s$^2$. }
\label{KH-Steps}
\end{figure}

An example of such a measurement is illustrated in Fig.~\ref{KH-Steps} at constant temperature above the transition to turbulence. Here the rotation $\Omega$ is increased at constant slow rate across the region where the Kelvin-Helmholtz instability of the AB interface develops. The $\Omega$ increase is performed sufficiently slowly so that no dynamic effects are visible. The first critical event we denote with $\Omega_{\rm c}$. During further increase of the rotation drive the instability occurs recurrently every time when the counterflow at the outer sample perimeter reaches the critical value $v_{\rm c} = \Omega_{\rm c} R_{\rm eff}$, where $r = R_{\rm eff}$ is the radial value at which the injected vorticity breaks through the AB interface (Sec.~\ref{Injection}). Simultaneously a central cluster of rectilinear vortex lines starts to develop (Fig.~\ref{RotSupFluidStates} right).

The parameters controlling the instability are roughly identical in the upper and lower halves of the container. The first instability event (with vortex-free flow in the two B-phase sections) is registered at almost exactly the same value of $\Omega$ in both detector coils, as seen in Fig.~\ref{KH-Steps}. Nevertheless, the instabilities of the two AB interfaces in Fig.~\ref{ExpSetUp} are not coupled, but occur independently. Thus the subsequent instabilities in Fig.~\ref{KH-Steps} are not synchronous, since the number of vortices injected in the B-phase sections at each event varies randomly. Also in the temperature regime where the transition from regular to turbulent vortex expansion takes place, it is quite possible to have a regular event in one half and a turbulent event in the second half. As the two $^3$He-B samples are independent, one can have, for instance, vortex-free flow in the upper B-phase section $(N=0)$ and roughly the equilibrium number of vortices in both the A-phase and the lower B-phase sections $(N=N_{\rm eq})$.

As $\Omega_{\rm c}$ is almost equal for both halves of the container, the first critical event can be triggered with a small step increase in $\Omega$ at both AB interfaces simultaneously. This triggers the injection of vortex loops into both $^3$He-B samples and their dynamic evolution towards the respective final states is then recorded independently. This is our normal measuring procedure when searching for the transition between regular and turbulent vortex expansion. The top and bottom halves of the container give thus similar information, one half would be sufficient to identify the transition, but two halves gives better statistics. For the initial diagnostics, to understand the critical characteristics of the different sources of vortex formation in a new sample tube, the two independent NMR detection systems are most useful. Four different quartz tubes were examined in this setup. Their properties turned out to vary substantially with respect to vortex formation in the absence of the barrier field and the AB interfaces. This means that at present the cleanliness and the properties of the fused quartz surfaces are not in reliable control. The fourth and last sample tube was by far the best, with the highest overall critical velocity. Most of the results of this report were measured with this tube.
\vspace{-5mm}

\subsection{NMR Measurement}\label{NMR}

The classification of the vortex expansion events is based on a measurement of the number of rectilinear vortex lines in the final state. The vortex line number $N$ can be accurately determined, if needed, from the linear reduction in the CF peak height as a function of $N$.  It turns out that precise measurements of $N$ are not needed. Because of the narrow width of the transition regime between regular and turbulent vortex dynamics at high rotation velocities, we need to distinguish only between two states, one with a small vortex cluster $(N \ll N_{\rm eq})$ at temperatures above the transition and one with the equilibrium vortex state $(N = N_{\rm eq})$ at temperatures below the transition. In Fig.~\ref{ExpSetUp} the two line shapes with very different appearance exemplify these two cases.

\begin{figure}[tp]
\begin{center}
\leavevmode
\includegraphics[width=0.9\linewidth]{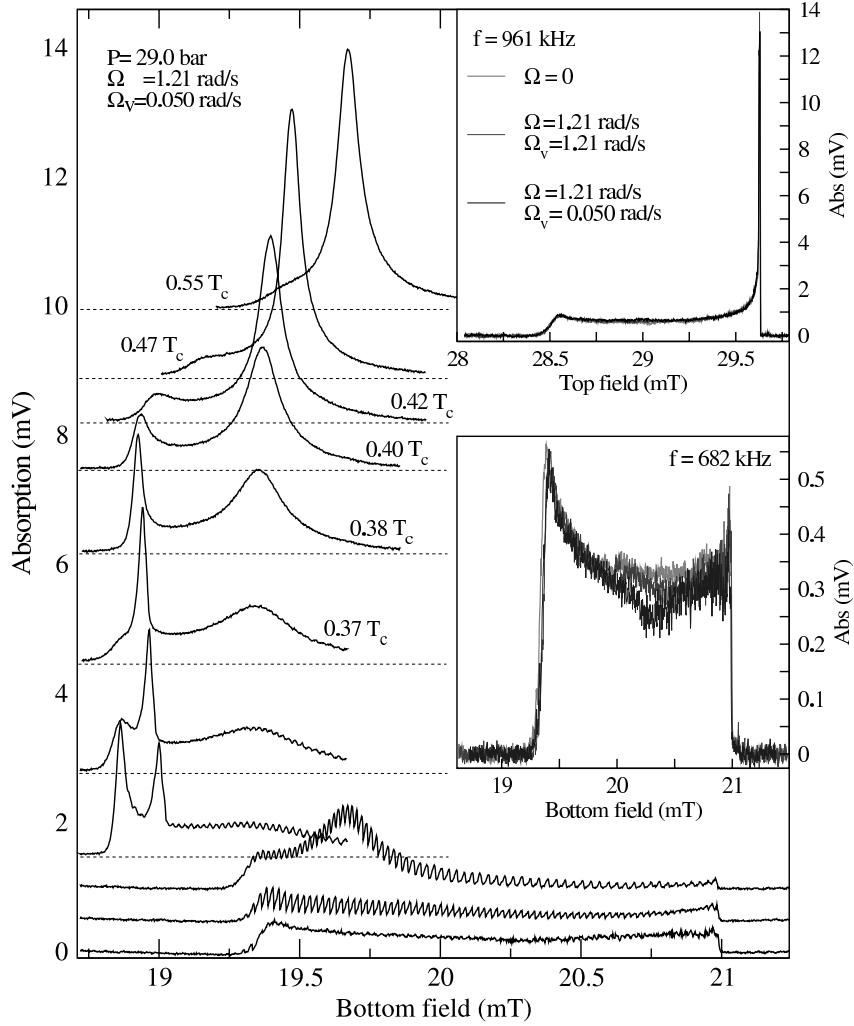}
\caption{Spectra of $^3$He-B during continuous cool down at constant rotation. {\it (Main panel)} With decreasing temperature the CF peak is shifted further from the Larmor value (at 21.0\,mT) and its height is reduced, since the total absorption is proportional to the static susceptibility $\chi_{\rm B}(T,P,H)$. Below $0.4\,T_{\rm c}$ the textural orienting interaction from the CF vanishes rapidly with decreasing temperature, as shown by this series of CF line shapes recorded with the bottom spectrometer at constant $\Omega$ and $N= \pi R^2 \, 2\Omega_{\rm v}/\kappa \lesssim 40\,$vortex lines. All spectra extend with finite (but small) absorption up to the Larmor value, but the line shapes above the three bottom examples have been truncated. {\it (Inserts)} Spectra from the top and bottom spectrometers measured simultaneously at the lowest temperature. In both inserts three different situations have been compared: a) sample not rotating $(\Omega = 0)$, b) sample filled with the equilibrium number of vortex lines, and c) continuous cool-down in rotation at constant $\Omega $ with a small stable vortex cluster. }
\label{NMR-CF-Spectra}
\end{center}
\vspace{-6mm}
\end{figure}

The large difference between the two line shapes in Fig.~\ref{ExpSetUp}  is caused by the textural orienting influence of the macroscopic counterflow. In rotating experiments the textural interaction can be conveniently manipulated with the rotation velocity $\Omega$. The velocity of the counterflow $v(N, \Omega)$ is given by
\begin{equation}
v(N,\Omega) = v_{\rm n} - v_{\rm s} = (\Omega - \Omega_{\rm v})r = \Omega r  - \kappa N/(2\pi r)\;
\label{CF-Velocity}
\end{equation}
within the vortex-free annulus, which surrounds the central vortex cluster
in Fig.~\ref{RotSupFluidStates} (right), {\it i.e.} in the region $R
\sqrt{\Omega_{\rm v}/\Omega} < r < R$. Here the rotation velocity
$\Omega_{\rm v}(N)$ is an experimentally useful quantity which denotes the
rotation at which the cluster of $N$ lines fills the whole cross section of the sample, {\it
  ie.} the rotation velocity at which a given number of lines $N$ is in the
equilibrium state. No changes in the vortex-line number occur if one sweeps $\Omega$ between the limits $\Omega_{\rm v}(N) \leq \Omega < \Omega_{\rm c}$, where $\Omega_{\rm c}(T,P)$ is the relevant critical velocity, at which more vortices start to form. By changing $\Omega$, one scans the magnitude of the textural interaction term\cite{DensityAnisotropy}
\begin{equation}
 F_{\rm hv} = -\frac{1}{2}\, \delta \rho_{\rm s} ({\hat {\bf l}} \cdot {\bf v})^2\,,
\label{CF-Interaction}
\end{equation}
where the superfluid density anisotropy $\delta \rho_{\rm s} = \rho_{{\rm
  s}\perp} - \rho_{{\rm s}\parallel}$. The applied magnetic field
induces an orbital momentum in the direction $l_i = R_{\alpha i} {\hat
  h}_{\alpha}$ where the unit vector ${\hat {\bf h}} = {\bf H}/H$ specifies
the direction of the applied field ${\bf H}$, and $R({\hat {\bf n}},
\theta)$ is a rotation matrix in the B-phase order parameter (which minimizes the dipolar spin-orbit interaction in the equilibrium state). Since $\delta \rho_{\rm s} > 0$, to minimize the free energy $F_{\rm hv}$ in the equilibrium configuration of the texture,
the azimuthal CF velocity ${\bf v}$ attempts to orient the orbital
anisotropy axis ${\hat {\bf l}}$ parallel to itself. Since the magnetic
field induced distortion is of axial symmetry, in first order the
anisotropy $\delta \rho_{\rm s} \propto H^2$. In addition $F_{\rm hv}
\propto \Omega^2$ (assuming $N \approx 0$) and thus with increasing
rotation the orienting effect from the CF becomes comparable sequentially
with other textural interactions. In the axially symmetric case (when the
magnetic field is oriented along the symmetry axis of the sample cylinder,
${\hat {\bf h}} \parallel {\hat {\bf z}} \parallel {\mathbf \Omega}$) this
leads to a sequence of different axially symmetric textures, which are
generally known to be of ``flare-out'' configuration.\cite{CF-Textures}

The density anisotropy $\delta \rho_{\rm s}(T,P)$ is only weakly temperature dependent, when $T > 0.4\, T_{\rm c}$. This is the regime where all NMR measurements on vortices in rotating $^3$He-B have so far been performed and where the measuring techniques are well established. Below $0.4\, T_{\rm c}$ however, $\delta \rho_{\rm s}$ drops rapidly with decreasing temperature.\cite{DensityAnisotropy} The evolution of the NMR absorption line shape during cooling (in otherwise constant conditions) is illustrated in the main panel of Fig.~\ref{NMR-CF-Spectra}: With decreasing temperature the peak heights rapidly drop, the line shapes undergo changes owing to textural transitions, standing spin-wave resonances grow in amplitude and decorate the line shape, and finally the orienting influence of the CF on the texture vanishes, the spin wave amplitudes die away, and the NMR shifts saturate below $0.35\,T_{\rm c}$ (bottom spectrum). The textural transitions as a function of the magnitude of the CF interaction (\ref{CF-Interaction}) have been studied at temperatures above $0.4\, T_{\rm c}$ in Ref.~\onlinecite{CF-Textures}. Broadly speaking the transitions in Fig.~\ref{NMR-CF-Spectra} resemble those which occur at higher temperatures while $\Omega$ is reduced at constant temperature. An example is the second CF peak at maximum possible NMR shift which starts to grow in amplitude at $T \lesssim 0.42\, T_{\rm c}$. At still lower temperatures this second peak breaks into a twin peak structure while the main CF peak at smaller NMR shift is already declining in amplitude. The doubling of the peak at maximum possible NMR shift  has not been observed before and no  explanation exists for it. Finally below $0.35\, T_{\rm c}$ the effect of the counterflow is not visible in the line shape and the final example on the bottom of Fig.~\ref{NMR-CF-Spectra} is similar in appearance to the equilibrium vortex state spectrum in Fig.~\ref{ExpSetUp}.

The two inserts to Fig.~\ref{NMR-CF-Spectra} demonstrate that at the lowest temperature the NMR spectrum is completely insensitive to rotation: The line shapes are identical irrespective whether the sample (a) is not rotating at all ($\Omega = 0$), (b) rotating in the equilibrium vortex state $(N=N_{\rm eq}$), (c) or with vortex-free flow around a small vortex cluster ($\Omega_{\rm v} = 0.050\,$rad/s).  The NMR polarization field of the top spectrometer is 40\,\% higher than that of the bottom spectrometer. In the upper insert of Fig.~\ref{NMR-CF-Spectra} the textural magnetic healing length\cite{Hakonen} is of order $\xi_{\rm H} \sim \frac{1}{3} \, R$ and the line shape is that of a regular flare-out texture. In the lower insert at lower NMR field $\xi_{\rm H} \propto 1/H$ is larger and the line shape is more distorted. The spectra from the top and bottom spectrometers in these two inserts have been measured simultaneously in zero barrier field ({\it i.e.} in the absence of an A-phase layer) at the lowest temperature to which the sample cools. This temperature is limited by the residual heat leak into the $^3$He-B sample and its thermal contact to the refrigerator.  Owing to the saturation of the NMR shifts, and to a lesser degree the uncertainty in the value of the Larmor edge ({\it i.e.} the line width from the residual magnetic field inhomogeneity of the NMR polarization magnets, FWHH $\sim 0.8\,$kHz), with the presently available techniques the lowest temperature can only be assigned an upper limit $T < 0.35\, T_{\rm c}$.

An interesting further feature in the line shapes of Fig.~\ref{NMR-CF-Spectra} are spin-wave resonances in the two signals just above that on the very bottom. In contrast to the local oscillator model, which yields the overall line shape of the flare-out textures, spin-wave absorption modes represent coherent spin precession over global regions with a smoothly varying texture.\cite{Hakonen} The magnetic healing length of the texture $\xi_{\rm H}$ grows with decreasing temperature and becomes finally comparable to the sample radius $R$. This causes gradients in textural orientations to become more and more gradual. Thus the spin-wave resonances grow in prominence with decreasing temperature. As seen in Fig.~\ref{NMR-CF-Spectra}, both the intensity and the width of successive spin-wave resonances are approximately constant across the entire absorption regime of the signal. In fact, theory predicts that their absorption intensity grows towards low temperatures and finally saturates, similar to NMR shifts. However, in the signal corresponding to the lowest temperature in Fig.~\ref{NMR-CF-Spectra} spin-wave resonances are not visible. It is believed that here, owing to the absence of dissipation and rigidity of the order parameter texture, spin-wave resonances are not displayed by our measurement.

The most convenient thermometer in these measurements is the horizontal shift of the CF peak from the Larmor value. The shift is calibrated in a slow continuous warmup, which is repeated at varying warming rates, against two external thermometers. These are rigidly fixed on the nuclear cooling stage: a capacitive $^3$He melting pressure gauge and a pulsed NMR thermometer with a platinum wire brush as sample.  Our temperature calibration of the NMR shift is generally in agreement with those reported in Refs.~\onlinecite{Ahonen} and \onlinecite{Hakonen}. However, below 0.7\,mK  the external thermometers lose contact with the $^3$He-B sample and a smooth extrapolation of the NMR shift has to be used as temperature scale. Finally, as seen in Fig.~\ref{NMR-CF-Spectra}  below $\sim 0.35\,T_{\rm c}$, the NMR shifts saturate and become useless for thermometry.

To summarize, in the present type of NMR measurements the experimentally vital information is obtained via the image which the NMR provides about the order parameter texture, primarily through the change in the magnitude of the counterflow interaction in Eq.~(\ref{CF-Interaction}) as a function of the number of rectilinear vortex lines $N$ at constant $\Omega$. These techniques work well at temperatures above $ 0.4\, T_{\rm c}$, so that single-vortex resolution is achieved above $ 0.7\, T_{\rm c}$. However below $ 0.35\, T_{\rm c}$ the texture-based NMR techniques break down. This is most unfortunate for further efforts to study the properties of turbulence in the zero temperature limit. Clearly $T \geq 0.35\, T_{\rm c}$ is not yet the regime of ballistic quasiparticle motion and vanishingly small mutual friction. At present it is not clear if other NMR techniques can be developed for $^3$He-B to measure the number of rectilinear vortex lines or the vortex density at the lowest temperatures.
\vspace{-5mm}

\subsection{Vortex Loop Injection}\label{Injection}

\begin{figure}[t]
\centerline{\includegraphics[width=0.9\linewidth]{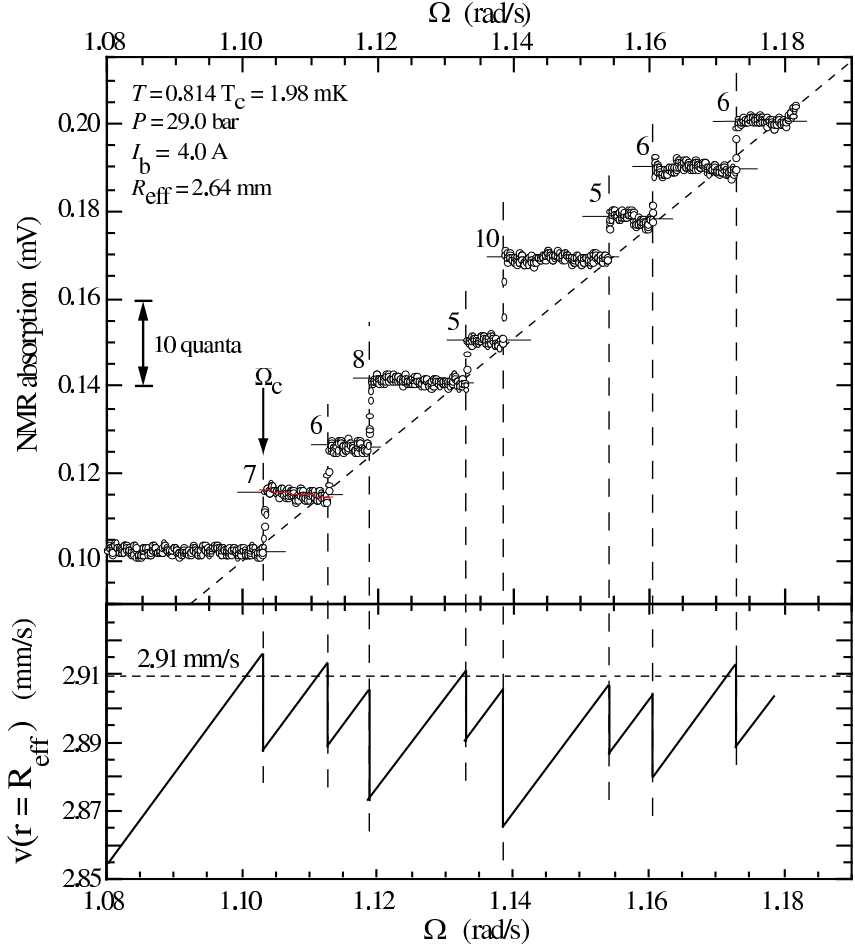}}
\medskip
\caption{A sequence of KH instability events with increasing $\Omega$. The absorption amplitude close to the Larmor value has been recorded with the bottom spectrometer. The sloping dashed line is a fit through the instability points and is proportional to $N = \pi R_{\rm eff}^2 \, 2 (\Omega - \Omega_{\rm c})/\kappa$.  {\it (Bottom)} The discontinuous CF velocity as a function of   $\Omega$ at $r = R_{\rm eff}$: $v = \mid v_{\rm sB} - v_{\rm n} \mid = \Omega  R_{\rm eff} - \kappa N/(2\pi R_{\rm eff})$. The horizontal dashed line is equivalent to the sloping dashed line in the upper panel and defines the mean critical velocity $v_{\rm c}$.}
\label{KH-HiPrecisionSteps}
\end{figure}

The classic Kelvin-Helmholtz (KH) instability is a generic phenomenon of fluid dynamics which occurs at the interface between two inviscid fluid layers, which move with respect to each other in a state of relative shear flow. If the difference in tangential velocities is small, the interface remains calm, but at some critical value interfacial waves are formed. The AB interface instability is its first superfluid example.\cite{KH}

$^3$He-A and B behave very differently in rotation: The A phase has a low
critical velocity and remains close to the equilibrium vortex state at all
times, while the B phase is vortex-free before any instabilities have
occurred. To maintain this situation, a new structure of vorticity is
formed on the A-phase side of the AB interface, a surface vortex layer. It
is confined to the interface and gives rise to the tangential velocity
difference across the phase boundary.\cite{AB-VortexLayer} The instability
manifests itself as a sudden event in which some vortices from the
surface layer break through the interface and start to
evolve in the vortex-free B-phase counterflow.

The critical CF velocities across the interface follow the general formula\cite{KH-Theory}
\begin{equation}
{1 \over 2}\, \rho_{\rm sA} (v_{\rm sA} - v_{\rm n})^2 + {1 \over 2}\, \rho_{\rm sB}
(v_{\rm sB} - v_{\rm n})^2 = \sqrt{\sigma_{\rm AB} F}\;.
\label{KH-CritVelocity}
\end{equation}
where the magnetic restoring force on the AB interface is $F = {1 \over
  2}\, (\chi_{\rm A} - \chi_{\rm B}) \, \nabla (H_{\rm b}^2)$ at $H_{\rm b}
= H_{\rm AB}$. Here $\chi_{\rm A}(P)$ and $\chi_{\rm B}(T,P,H)$ are the susceptibilities of the A and B phases, $\sigma_{\rm AB}(T,P)$ the surface tension of the AB interface, and $H_{\rm AB}(T,P)$ the critical magnetic field of the AB transition. In our experiments $v_{\rm sA} \approx v_{\rm n}$ and $v_{\rm sB} - v_{\rm n} = \Omega_{\rm c} R_{\rm eff}$, where $\Omega_{\rm c}$ is the first critical rotation velocity of the KH instability and $R_{\rm eff}$ the effective radial value for the instability. The wave vector of the mode of interface waves, which causes the instability, is $k = \sqrt{F/\sigma}$. Measurements in widely
different conditions agree with Eq.~(\ref{KH-CritVelocity}) without adjusted or fitted  parameters.\cite{KH,ToBePublished}

\begin{figure}[t]
\centerline{\includegraphics[width=0.8\linewidth]{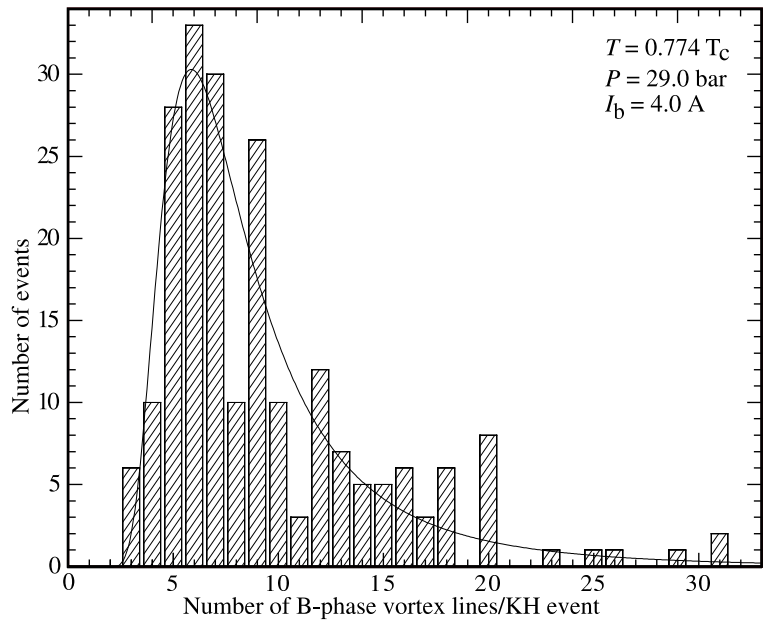}}
\medskip
\caption{Histogram of 214 measured shear-flow instability events of the AB
interface, plotted as a function of the number of rectilinear B-phase vortices $\Delta N$
formed per event. The smooth curve is a guide for the eye of a differential probability distribution.} \label{StepDistribution}
\end{figure}

A repetition of Fig.~\ref{KH-Steps} at higher signal-to-noise resolution is shown in Fig.~\ref{KH-HiPrecisionSteps} for the bottom section of the sample. It illustrates the experimental consequences from Eq.~(\ref{KH-CritVelocity}), the signatures of the KH instability. A sequence of instability events has been recorded here in slowly increasing rotation at high temperatures where the number of vortex loops injected in the B phase equals the number of rectilinear vortex lines which are formed after each instability event. The signal monitored in the top panel is the absorption close to the Larmor edge where the step-like increase measures the number of new rectilinear vortex lines which are added to the lower B phase section, {\it i.e.} the vertical discontinuous step height is proportional to the increase in vortex number. The simplest technique to assign the correspondence between step height and vortex number $\Delta N$ is to perform a least squares fit with the constraint that the values for $\Delta N$ have to be integral numbers. The order of magnitude for the step height signal is $2\,\mu$V/vortex line. This figure refers to the voltage measured at the output of our cryogenic MESFET preamplifier (with a gain $\sim 10$) which follows the superconducting tank circuit. With this fit we get the vortex line numbers which are denoted in Fig.~\ref{KH-HiPrecisionSteps} next to each step.

The location of the instability events on the $\Omega$ axis has to fulfill the requirement that the critical KH velocity is constant, {\it i.e.} the corner points in the staircase pattern fall on one line. This gives the dashed linear fit with the slope $1.27\,\mu$V/(rad/s). Combining the two numbers so far, we get $\Delta N/\Delta \Omega = 671\,$lines/(rad/s). This means that when $\Delta N$ new rectilinear lines are added to the central vortex cluster, the counterflow velocity along the effective perimeter decreases by $\Delta \Omega \cdot R_{\rm eff} = \kappa \, \Delta N/(2\pi R_{\rm eff})$. From this requirement we obtain the effective radial value $R_{\rm eff} \approx 2.6\,$mm for the instability, which should be compared with the sample radius $R = 3.0\,$mm.

A lot of similar examples confirm the above value for the effective radius $R_{\rm eff} < R$. Also more elaborate measurements, which measure the number of vortex lines formed in each instability event with a calibrating procedure, give the same value of $R_{\rm eff}$ in the above regime of $\Omega, T, P$.  A result from this latter type of precision measurements was reported in Ref.~\onlinecite{AB-VortexLayer}. What then is the explanation for $R_{\rm eff}$?  The vortices which create the staircase pattern in Fig.~\ref{KH-HiPrecisionSteps} are detected by the pick-up coil in the bottom of the sample tube 29\,mm below the AB interface. The flight time for the vortices from the interface to the top edge of the coil is 5\,s. At high temperatures the expansion of the vortices is deterministic. The fixed flight time introduces a 5\,s delay before the information about the events at the interface starts to build up in the pick-up coil. In Fig.~\ref{KH-HiPrecisionSteps} the rotation drive is a linear ramp with the  acceleration rate $d\Omega/dt = 2.5 \cdot 10^{-4}\,$rad/s$^2$. This means that every 6\,s the counterflow velocity increases at the effective perimeter by the equivalent of one circulation quantum. Comparing these two numbers, we interpret the staircase pattern to arise from the events at the interface and not to be blurred by the motion of the vortices along the rotating column. In this case $R_{\rm eff}$ is the radial distance at which the KH instability occurs and the vorticity is injected across the AB  interface from the A phase into the B phase.

The fact that we measure a definite value for $R_{\rm eff}$ allows us to make a guess of the initial configuration of the injected vortex loops immediately after injection, in the situation from where their expansion in the vortex-free B-phase counterflow starts. This would be a loop of length $\gtrsim R-R_{\rm eff} \approx 0.4\,$mm whose one end sticks out of the AB interface at $r=R_{\rm eff}$ and the second ends on the cylindrical side wall at $r=R$ in the B phase section.

The second important question about the injection is the number of such loops. In Fig.~\ref{KH-HiPrecisionSteps} we have a small sample of eight instability events, with the number of vortex lines $\Delta N$ formed per event marked next to each event.  It is seen that  $\Delta N$ is a small random number and on an average $\Delta N \approx 7$. In Fig.~\ref{StepDistribution} a histogram  of $\Delta N$ for a large sample of 214 events is shown.  Such a probability distribution can only be measured at high temperatures with regular vortex motion. It has not been carefully measured as a function of the external variables $T, \Omega, P$, but there are no indications that it would be rapidly changing in the conditions used for vortex loop injection in this work. Therefore we assume that the distribution behaves in a continuous manner across the transition from regular to turbulent dynamics and is of similar shape in the turbulent regime.

The average of the probability distribution in Fig.~\ref{StepDistribution} is 11 vortex lines per event. This agrees with what one would expect for the number of A-phase vortex quanta in the surface vortex layer which cover one trough, or half a wave length, in the wave pattern of the interface ripplons according to Eq.~(\protect\ref{KH-CritVelocity}). In the A-phase section, assuming solid-body rotation with zero critical velocity, there are $N \approx \pi R_{\rm eff}^2 \, 2\Omega_{\rm c}/\kappa$ single-quantum vortices ($\kappa \! = \! h/2m_3$) covering the AB interface which all flare out radially perpendicular to the cylindrical wall. Measured along the effective perimeter there are thus $\Omega_{\rm c}R_{\rm eff}/\kappa$ circulation quanta per unit length and thus in one corrugation $\Delta N  \approx
({\pi v_{\rm c}})/({k\kappa})$, where $v_{\rm c} = \Omega_{\rm c}R_{\rm eff}$. From Eq.~(\protect\ref{KH-CritVelocity}) this is seen to be $\Delta N \approx  (2\pi \sigma)/(\kappa v_{\rm c}\rho_{\rm sB})$, which gives 9 vortex lines in fair agreement with the measured average in Fig.~\ref{StepDistribution} ($v_{\rm c} =  0.39$\,cm/s at $0.77\,T_{\rm c}$, 29.0\,bar, and $I_{\rm b} = 8.0\,$A).

\begin{figure}[t]
\centerline{\includegraphics[width=1.0\linewidth]{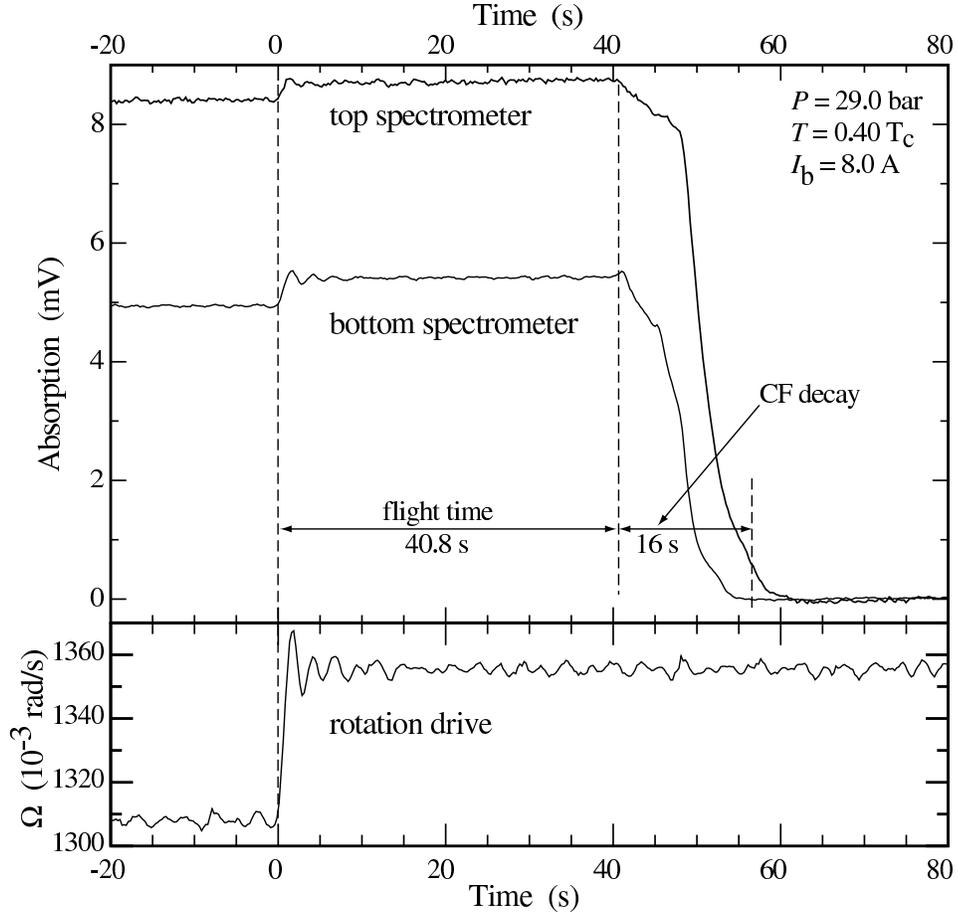}}
\medskip
\caption{Response of the CF peak height to a triggered KH injection event
  in the turbulent temperature regime. {\it (Bottom)} The instability is
  triggered by increasing $\Omega$ from 1.31 to 1.36\,rad/s in 2\,s. With
  vortex-free flow in both B-phase sections, the instability at both AB
  interfaces falls between these two limits. {\it (Top)} The trigger step increase
  $\Delta \Omega$ is instantaneously registered as an increase in
  the CF peak height. A rapid decay of the peak height starts when the
  front of propagating vorticity reaches the closest end of the detector
  coil and approaches zero when the front leaves the far end of the coil.
  This signal decay corresponds to the removal of the large-scale azimuthal
  counterflow just behind the propagating front. The NMR polarization
  fields are maintained here at constant value at the location of the
  maximum of the CF peak of the initial state ({\it e.g.} at 19.37\,mT for
  the bottom spectrometer in Fig.~\ref{ExpSetUp}).  }
\label{InjectionTurbulence}
\end{figure}

As noted in the context of Fig.~\ref{KH-Steps}, the first critical event, before any vortices have been injected, takes place almost simultaneously in both B-phase sections. Therefore, when the critical rotation velocity $\Omega_{\rm c}$ is known, the two events can be triggered together with a small step increase $\Delta \Omega = 0.050\,$rad/s in the manner shown in Fig.~\ref{InjectionTurbulence}. This procedure allows to measure the flight time for the vorticity to propagate from the AB interface to the detector coil. Moreover, it allows to set up the NMR measurement so that transient signals can be captured efficiently when the vorticity front travels through the detector coils. In contrast to Fig.~\ref{KH-Steps}, in Fig.~\ref{InjectionTurbulence} the evolution of the CF peak height after injection has been monitored at low temperatures where the expansion of the injected vortex loops is turbulent.

What happens at injection? What is the configuration in which the vorticity
expands along a rotating superfluid column? The answers to these questions
are still discussed; we are not going to delve into details here. Clearly
the first turbulent proliferation of vorticity takes place immediately at
or after injection. If the rotating vortex-free column is infinitely long,
it is unrealistic to expect that turbulence would fill at some point
of the process the entire column. Rather it could be expected to expand
into the vortex-free section in a form which after some time travels along
the column in a time-invariant average configuration. Since the flight time
of the vorticity grows as $\tau_{\rm F} = d/(\alpha \Omega R)$ with
decreasing friction $\alpha$, the propagating front moves slower with
decreasing temperature and has more time to settle down in such a
configuration. Thus at $0.40\,T_{\rm c}$ the transient signals do not
appear to depend on the distance $d$. This conclusion is reached from
monitoring the transient signals, which the propagating front induces in
the two detector coils when turbulence is started in random locations along
the sample with a neutron capture process (in zero barrier field and no A-phase).\cite{NeutronTurbulence} With this picture in mind, we take a second look at the timing in Fig.~\ref{InjectionTurbulence}.

The split-half detector coils, whose axes are oriented transverse to the sample column in the setup of Fig.~\ref{ExpSetUp}, have a sharp cutoff in sensitivity at the edge of the coil. Thus the flight time from the AB interface to the detector coil is bracketed between the instantaneous increase in CF peak height from the trigger and the arrival of the vorticity front at the closer end of the detector coil (from where the decay of the CF signal starts in Fig.~\ref{InjectionTurbulence}). In Fig.~\ref{ExpSetUp} this flight time of 41\,s corresponds to a distance $d \approx 34\,$mm (bottom section of sample). The front is expected to travel through the 10\,mm long detector coil during the next $10/34\times 41\,{\rm s}= 12\,$s. From Fig.~\ref{InjectionTurbulence} we note that the CF peak height of the bottom spectrometer vanishes in $\sim 14\,$s. Comparing these two numbers, this means that the macroscopic azimuthally flowing counterflow is removed immediately behind the traveling front of vorticity, {\it i.e.} the vorticity behind the first front has the equilibrium polarization of solid body rotation in the direction $\hat {\mathbf z}$ of the column. Other conclusions about the structure of the front in the turbulent temperature regime can be drawn from the transient signal measured at the Larmor edge, {\it i.e.} by monitoring the build up of the vorticity instead of the CF decay.\cite{ToBePublished}

Figs.~\ref{KH-HiPrecisionSteps} and \ref{InjectionTurbulence} thus characterize different aspects of how the vorticity expands in the rotating column at high and low temperatures. The most fundamental difference is in the final states: In Fig.~\ref{KH-HiPrecisionSteps} only a small number of rectilinear vortex lines is created in each sequential KH event, while in Fig.~\ref{InjectionTurbulence} the first event fills the vortex-free B-phase section with the equilibrium number of vortex lines. We have measured extensively the first critical velocity $\Omega_{\rm c}$ of the KH instability as a function of $T, I_{\rm b}, P$. These measurements give in the $(\Omega,T)$ plane continuous trajectories,\cite{KH} in which the transition to turbulent vortex dynamics is not visible. There is no indication that the injection process itself would influence which final state is formed, when the temperature is well above or below the transition. Rather, all measurements suggest that the KH instability is a predictable injection mechanism with continuous properties and with a reasonably narrow distribution in the number of injected vortex loops.

The measurement of flight time $\tau_{\rm F}$ in Fig.~\ref{InjectionTurbulence} gives the propagation velocity of the vorticity along the rotating column: $v_{\rm Lz} = d/\tau_{\rm F}$. It turns out to amount to $v_{\rm Lz} = \alpha \Omega R$, since this expression gives $\alpha$ values in good agreement with the mutual friction measurements in Ref.~\onlinecite{Bevan}. This is the value expected for the end point of a single vortex spiralling down along the vertical container wall in the vortex-free state. The measurement can thus be used to determine the dissipative mutual friction coefficient $\alpha(T,P)$. Surprisingly the single-vortex value for $v_{\rm Lz}$ applies for the propagation velocity both in the regular and turbulent regimes of propagation, {\it i.e.} the velocity of the front of expanding vorticity with the retreating vortex-free flow displays no anomaly at the transition.\cite{FlightTime}

 \section{TRANSITION TO TURBULENCE} \label{TurbTransition}

When measured with KH injection, the transition from regular to turbulent dynamics is found to be velocity independent and centered at 0.52 -- 0.59\,$T_{\rm c}$, depending on pressure.\cite{TurbPhaseDiagram} It coincides with the condition $q = \alpha/(1-\alpha^{\prime}) \approx 1$. However, the transition is liquid pressure dependent, moving with increasing pressure to higher temperatures on the normalized $T/T_{\rm c}$ scale and also to higher $q$. Thus the transition does not take place at exactly the same $q$ value at all pressures and the situation is more complex than the simple theory in Sec.~\ref{Analysis} portrays. With KH injection the transition has a narrow width of $0.06\,T_{\rm c}$ at the three measured pressures of 10.2, 29.0, and 34\,bar. In this section we describe results from the transition region which shed more light on the KH injection mechanism.

\begin{figure}[t]
\centerline{\includegraphics[width=0.7\linewidth]{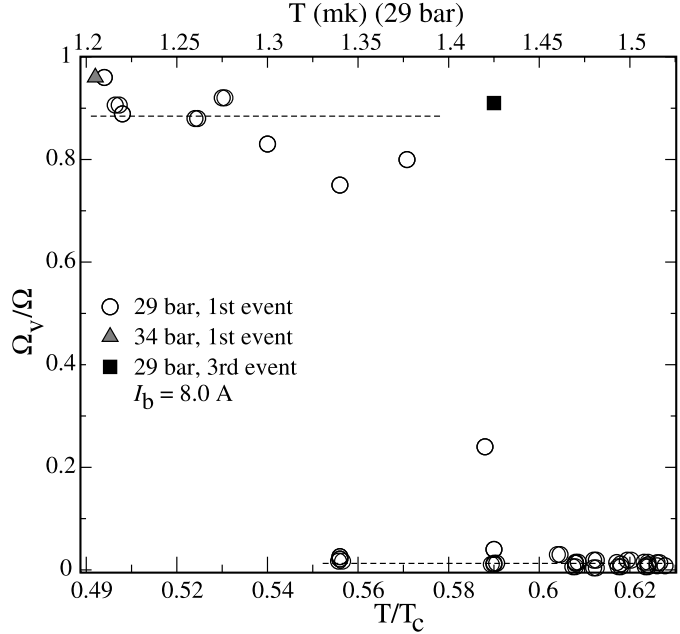}}
\medskip
\caption{Number of rectilinear vortex lines formed in a KH injection event as a function of temperature in the transition regime. The measured result on the vertical scale gives the fraction $\Omega_{\rm v}/\Omega$, the number of vortices present compared to the equilibrium number. This ratio is obtained using a calibration procedure similar to that explained in Ref.~\protect\onlinecite{NeutronTurbulence}. In most cases the result pertains to the first injection event with vortex-free flow in the B-phase section, but some cases have been included where the first or even the second sequential event gives only a small number (as in Fig.~\ref{StepDistribution}), until a later injection event produces a large number of vortices. The measurement demonstrates that the temperature width of the transition regime is narrow. Also the number of rectilinear vortex lines produced per injection event is either very small and the process is of the form shown in Fig.~\ref{KH-Steps}, or the final vortex number is close to  equilibrium and the process is as in Fig.~\ref{InjectionTurbulence}. In contrast,  intermediate vortex clusters are rare. } \label{TransitionRegime}
\end{figure}

A close up of the transition regime between regular and turbulent vortex motion as a function of temperature is shown in Fig.~\ref{TransitionRegime}. Here the number of rectilinear vortex lines has been measured and is expressed as a fraction of the equilibrium state, $\Omega_{\rm v}/\Omega$, where $\Omega$ is the critical rotation velocity $\Omega_{\rm c}$ of KH injection. The conspicuous feature is that intermediate values  $\Omega_{\rm v}/\Omega \sim 0.1$ -- 0.7 are prominently absent: the final state after injection includes either very few vortices or is close to equilibrium. In fact, the only intermediate values which have been measured are a few in the transition regime. This means that turbulence predominantly either switches on and fills the sample with close to the equilibrium number of vortices or it does not and the final state contains only the vortices injected by the KH instability. Therefore the transition regime is not characterized during cooling by a gradual increase in the value of $\Omega_{\rm v}/\Omega$, but by the fact that a growing fraction of the expansion processes after injection become turbulent and send the sample in the equilibrium vortex state.

The absence of intermediate values for the fraction $\Omega_{\rm v}/\Omega$ has an interesting explanation. It suggests that turbulence either switches on or does not depending on initial conditions, as was observed to be the case in neutron capture induced turbulence in Ref.~\onlinecite{NeutronTurbulence}. The initial conditions are dictated by the details of the injection process. Although the KH instability appears to be a reproducible injection technique, it contains a stochastic component, namely the number of injected vortex loops $\Delta N$ in Fig.~\ref{StepDistribution} and their exact configuration.  This would mean that it is not the expansion process, but the injection situation, which is most influential in starting turbulence: in the same conditions the expansion process might proceed towards only a few vortices or to the equilibrium number and it is the stochastic variability in KH injection
which produces the finite width of the transition.  These considerations are corroborated by ongoing numerical simulations of the appropriate experimental situation.\cite{Simulation}

\begin{figure}[t]
\centerline{\includegraphics[width=0.9\linewidth]{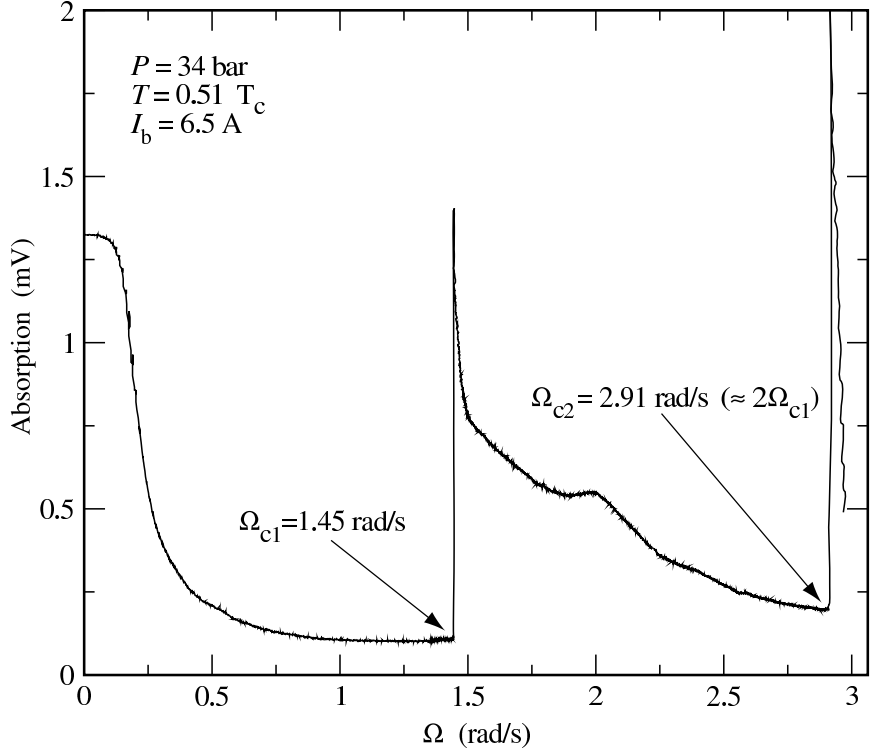}}
\medskip
\caption{Sequence of two turbulent transitions during a slow continuous acceleration of the rotation drive. The absorption response of the bottom spectrometer has been followed at constant polarization field just below the Larmor edge.} \label{TwoTurbTransitions}
\end{figure}

A further example of the transition features is illustrated in Fig.~\ref{TwoTurbTransitions}. It shows the NMR absorption response, measured at constant polarization field close to the Larmor edge, during a continuous slow increase of $\Omega$. The beginning of the trace $(\Omega < 1.4\,$rad/s) shows how the absorption is shifted from the Larmor region into the growing CF peak (compare to the spectra in Fig.~\ref{ExpSetUp}). Here it can be seen that the CF peak starts to develop only when $\Omega > 0.2\,$rad/s. Above 1\,rad/s the textural interaction from the CF has become the dominant contribution and from then on changes in the Larmor region are minor until the KH instability occurs at 1.45\,rad/s. The sharp peak here is a transient signal\cite{Turbulence} from a turbulent expansion process which fills the sample with almost the equilibrium number of vortices.

During further increase of $\Omega$ a central cluster of vortices with $\Omega_{\rm v} \approx 1.4\,$rad/s is formed, but otherwise a similar cycle of absorption behavior is repeated. Finally a second turbulent event takes place at exactly twice higher rotation velocity, as appropriate for the critical velocity of the KH instability according to Eq.~(\ref{CF-Velocity}). This latter turbulent event occurs in the annular vortex-free CF region which surrounds the central cluster in Fig.~\ref{RotSupFluidStates} (right). This example shows that (i) the central cluster formed from rectilinear vortex lines is stable in increasing rotation and (ii), since the second turbulent event occurs at the expected $\Omega$ value of the KH instability, the turbulent event evolves outside the cluster at large radii and CF velocities, so that the cluster is not involved in the process.

\begin{figure}[tp]
\centerline{\includegraphics[width=0.9\linewidth]{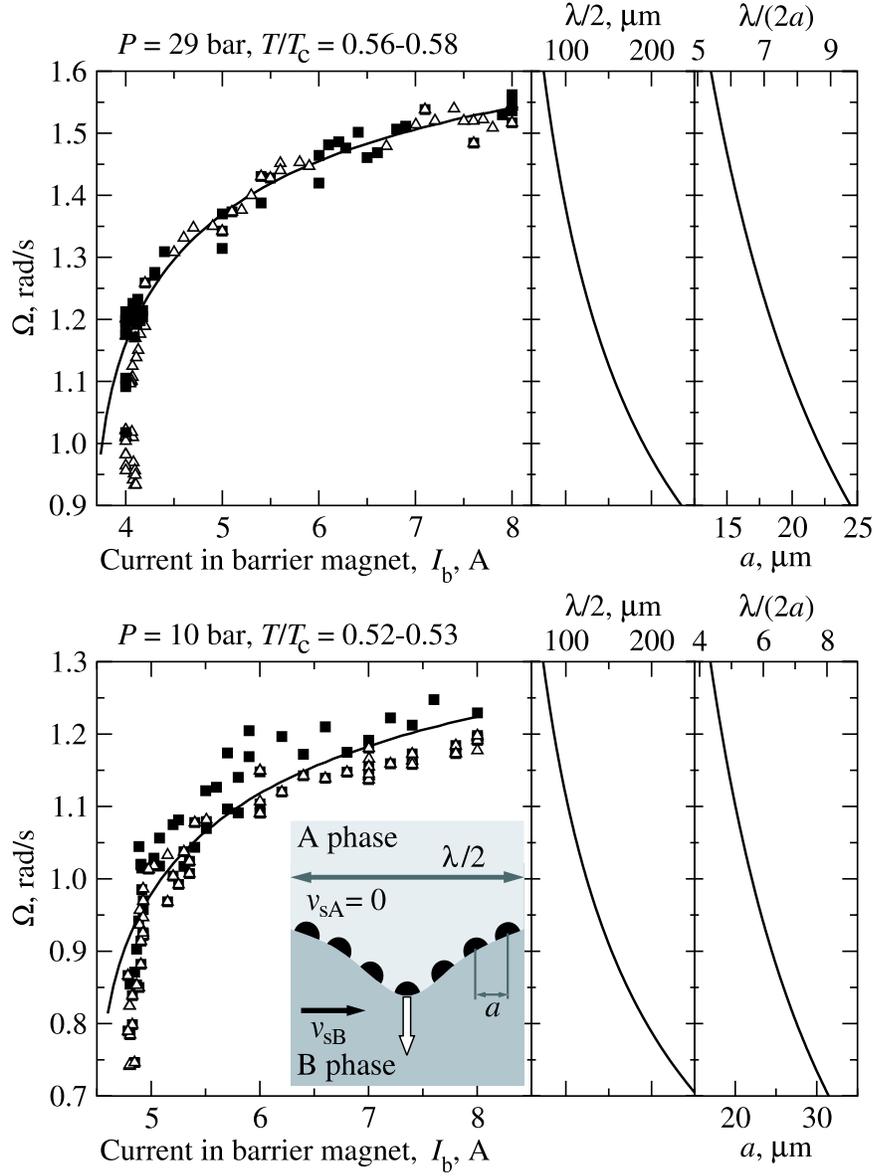}}
\medskip
\caption{Final states after KH injection in the temperature region of the transition from regular to turbulent vortex expansion, classified in (a) regular expansion events (open triangles), (b) turbulent events (filled squares), and (c) cases of one or a few regular events preceding a final turbulent event (triangle over a square). The curved solid lines represent Eq.~(\ref{KH-CritVelocity}) without adjustable parameters.\protect\cite{FootNote1} The insert in the bottom left panel is a cartoon of the interfacial wave on the AB phase boundary. The black half circles represent the A-phase vortex quanta covering the interface and flaring radially outwards ({\it i.e.} out of the plane of the paper).} \label{Transition}
\end{figure}

In Fig.~\ref{Transition} results are displayed on the final state after KH injection, as a function of CF velocity at constant temperature. The two plots examine the velocity dependence of the transition and here exactly in the temperature regime where the transition from regular to turbulent dynamics takes place. The data have been classified as (a) a regular expansion event if the final number of rectilinear vortex lines is in the range of the distribution in Fig.~\ref{StepDistribution},  as (b) turbulent if the final state contains close to the equilibrium number of rectilinear vortex lines, or as (c) mixed if one or a few regular events precede a final turbulent event. The data have been collected by moving as a function of the current $I_{\rm b}$ in the barrier magnet along an isotherm in the $(\Omega,T)$ plane. Since $T$ and $P$ are constant in this measurement, also the magnetic field at the interface is constant: $H = H_{\rm AB}$. On the left at low currents the curves start when the maximum field in the center of the  barrier solenoid at $z=0$ reaches $H_{\rm AB}$.  From here with increasing $I_{\rm b}$ the AB interfaces move apart along the solenoidal characteristic $H(I_{\rm b},z) = H_{\rm AB}$ and the gradient $\nabla H\mid_{H= H_{\rm AB}}$ increases. As a result, the KH critical velocity also moves to higher values, which is seen both from  Eq.~(\ref{KH-CritVelocity}) and in Fig.~\ref{Transition}.

Among the many different measuring runs which have been performed, the isotherms in the two panels of Fig.~\ref{Transition} happen to be almost exactly in the center of the transition regime at 10.2 and 29.0\,bar pressures. They were selected to illustrate the behavior in the transition regime. The results show no clear velocity dependence: The mix of different symbols in both panels appears random. This is also the general conclusion reached in Ref.~\onlinecite{TurbPhaseDiagram} when all data from the KH measurements is reviewed as a whole in the $\Omega$ range 0.8 -- 1.6\,rad/s.

The solid curves in the panels of Fig.~\ref{Transition} illustrate Eq.~(\ref{KH-CritVelocity}) and its predictions.  The panels in the middle show the behavior of the half length of the interface wave at the instability: $\frac{1}{2} \, \lambda = \pi\, \sqrt{\sigma_{\rm AB}/F}$. Inserting the critical KH velocity $\frac{1}{2} \, \rho_{\rm sB} \, v_{\rm c}^2 = \sqrt{\sigma_{\rm AB}\, F}$, we obtain
\begin{equation} \frac{1}{2} \, \lambda = 2\pi\, \frac {\sigma_{\rm AB}} {\rho_{\rm sB} R_{\rm eff}^2} \;  \frac{1} {\Omega_{\rm c}^2}
\label{lamda}
\end{equation}
and therefore the trough with the A-phase vortices, which break through the interface, gets rapidly smaller with increasing rotation. The inter-vortex distance $a$ in the surface vortex layer on the AB interface is determined by the tangential velocity difference: $a(v_{\rm sB} - v_{\rm sA}) \approx \kappa$ (see insert in panel on bottom left). The number of vortices in the trough can then be expressed as
\begin{equation} \Delta N \sim \frac{\lambda}{2a} = 2\pi\, \frac{ \sigma_{\rm AB}} {\kappa \rho_{\rm sB} R_{\rm eff}} \; \frac{1} {\Omega_{\rm c}} = 2\pi\, \frac{ \sigma_{\rm AB}} {\kappa^2 \rho_{\rm sB}}\; a
\label{DeltaN}
\end{equation}
and thus it also decreases with increasing rotation, as shown in the outermost panels on the right.

So far the probability distribution in Fig.~\ref{StepDistribution} has not been measured as a function of $\Omega$, to check the dependence $\Delta N \propto \Omega^{-1}$. However,it is instructive to consider the consequences from Eq.~(\ref{DeltaN}). Our measurements at 29.0\,bar and $I_{\rm b} =8.0\,$A as a function of temperature are a good example. First $\Omega_{\rm c}(T)$ was measured using a slowly accelerating rotation drive, as in Figs.~\ref{KH-Steps} and \ref{KH-HiPrecisionSteps}. $\Omega_{\rm c}(T)$ turned out to increase with decreasing temperature, curving smoothly towards saturation at $\Omega_{\rm c} \sim 1.6\,$rad/s below $ 0.6\,T_{\rm c}$. This agrees with Eq.~\eqref{KH-CritVelocity} and reflects the fact that at 8\,A the barrier field in the center of the solenoid exceeds $H_{\rm AB}(T \rightarrow 0)$. From Eq.~(\ref{DeltaN}) we see that if anything, then $\Delta N$ decreases towards low temperatures. The measurements were next repeated using the triggering method of Fig.~\ref{InjectionTurbulence} with $\Delta \Omega = 0.050\,$rad/s.  Such a measurement proceeded as seen in Fig.~\ref{InjectionTurbulence} and resulted below the transition regime $\lesssim 0.53 \, T_{\rm c}$ always in a turbulent event. In view of Eq.~(\ref{DeltaN}) this seems to mean that even with fewer injected vortex loops the first KH instability leads invariably to turbulence, or that even a small number of injected loops, such as $\Delta N \lesssim 3$ (Fig.~\ref{StepDistribution}), has to suffice to start turbulence. This we take as evidence for the importance of the Kelvin-wave instability\cite{Glaberson} of a single vortex as the initial  source for the proliferation of vorticity and as the mechanism which here starts the turbulence.

Finally we note in passing two features which are related to any future extension of these measurements: (i) With the present barrier magnet the trajectories of the KH instability do not extend much below or above the velocity interval shown in Fig.~\ref{Transition}. For lower velocities a barrier magnet with a weaker gradient is needed, for higher velocities it should be steeper (Eq.~\eqref{KH-CritVelocity}). (ii) We have performed measurements with KH injection also at zero pressure, to span the entire range in particle densities. However, at low temperatures below $\sim 0.5\,T_{\rm c}$ a new interfering feature appears, the continuous proliferation of vorticity when curved vortex lines are present and undergo the Kelvin-wave instability.\cite{Glaberson} It then becomes difficult to achieve vortex-free rotation to sufficiently high velocities, as required for the KH instability. This difficulty grows in prominence at low pressures, especially at zero pressure. The Kelvin-wave instability  starts to fill the sample at a seemingly constant rate with rectilinear vortex lines already at low rotation and proceeds until close to the equilibrium vortex state. The source of the problem are existing curved vortex lines, which at $\Omega =0$ are remanent filaments, left over from earlier rotations. These are difficult to avoid at low temperatures, since the last one or two remanent vortices take exceedingly long times to annihilate in our long sample tube in standstill $(\Omega =0)$. Rotation in the reverse direction is not helpful since the required velocities are of order $\sim 10^{-4}\,$rad/s and even then more often than not lead to a reorientation of the filament (a reversal in the orientation of $\bkappa$) and thereby via the Kelvin-wave instability to more vorticity. The long life time of vortex filaments at the lowest temperatures in zero applied flow we noted when the example of a single precessing vortex from Ref.~\onlinecite{VortexPrecession} was described in Sec.~\ref{Introduction}

\section{CONCLUSIONS}

We have described the techniques for investigating turbulence in $^3$He-B in the temperature range $T \gtrsim 0.4\,T_{\rm c}$.  Both the methods to generate turbulence and to record it non-invasively are different from what can be done in $^4$He-II. In the temperature region of these measurements the vortex mutual friction of $^3$He-B changes rapidly
and proves to control the character of the vortex dynamics. A sharp transition from regular vortex number conserving dynamics at high temperatures to turbulent dynamics at low temperatures is observed at $T \lesssim 0.6\,T_{\rm c}$. The best injection mechanism of vortex loops into vortex-free counterflow of $^3$He-B has been found to be the Kelvin-Helmholtz instability of the magnetically stabilized interface between $^3$He-A and $^3$He-B. In this process only the number of injected vortex loops is not in good control, which is believed to cause the observed finite transition width.

It is to be expected that further work on turbulence in $^3$He-B will allow
to place turbulence in $^4$He-II, the only previously known example of
superfluid turbulence, into a wider context. In contrast to $^4$He-II, where in most cases at
higher flow velocities one has to deal with the mutual-friction coupled
turbulence in both the superfluid and normal components, in $^3$He-B experiments the
normal component is practically always immobile. Thus a simpler example
of superfluid turbulence is available for studies here. The present NMR technique, which relies on the influence of the vortex-free counterflow on the order parameter
texture, stops working below $ 0.35\,T_{\rm c}$. At present it is not known if other NMR techniques can be developed to investigate turbulence in the zero temperature limit. \vspace{-3mm}

\section*{ACKNOWLEDGMENTS}

We thank Carlo Barenghi, Rob Blaauwgeers, Girgl Eska, Demosthenes Kivotides, Nikolai Kopnin,  Makoto Tsubota, and Grigory Volovik for valuable discussions and especially Ladislav Skrbek and Joe Vinen for a critical reading of this manuscript. This work has benefited from the EU-IHP ULTI-3 visitor program (HPRI-CT-1999-00050) and the ESF conference programs COSLAB and VORTEX.


\vspace{-5mm}
\begin{thebibliography}{99}

\bibitem{VinenNiemela} W.F. Vinen and J.J. Niemela, {\it J. Low Temp. Phys.} {\bf 128}, 167 (2002).

\bibitem{Johansen} E. Altshuler, T.H. Johansen, {\it Rev. Mod. Phys.} in print (2004).

\bibitem{CollectiveMotion}  M. Krusius, J.S. Korhonen, Y. Kondo, E.B. Sonin,  {\it Phys. Rev.} B {\bf 47}, 15113 (1993); {\it Europhys. Lett.} {\bf 22}, 125 (1993).

\bibitem{VortexPrecession} R.J. Zieve, Yu.M. Mukharsky, J.D. Close, J.C. Davis, R.E. Packard, {\it J. Low Temp. Phys.} {\bf 91}, 315 (1993); {\it Phys. Rev. Lett.} {\bf 68}, 1327 (1992).

\bibitem{Ruutu} V.M.H. Ruutu, \"{U}. Parts, J.H. Koivuniemi, N.B. Kopnin, M. Krusius,  {\it J. Low Temp. Phys.} \textbf{107}, 93 (1997).

\bibitem{AndreevReflection} S.N. Fisher, A.J. Hale, A.M. Gu\'{e}nault, G.R. Pickett, {\it Phys. Rev. Lett.} {\bf 86}, 244 (2001).

\bibitem{LineDensity} D.I Bradley, S.N. Fisher, A.M. Gu\'{e}nault, M.R. Lowe, G.R. Pickett, A. Rahm, R.C.V. Whitehead, {\it Phys. Rev. Lett.}, submitted.

\bibitem{KH} R. Blaauwgeers, V.B. Eltsov, G. Eska, A.P. Finne, R.P. Haley, M. Krusius, J.J. Ruohio, L. Skrbek, G.E. Volovik, {\it Phys. Rev. Lett.} {\bf 89}, 155301 (2002).

\bibitem{Turbulence} A.P. Finne, T. Araki, R. Blaauwgeers, V.B. Eltsov, N.B. Kopnin, M. Krusius, L. Skrbek, M. Tsubota, G.E. Volovik, {\it Nature} {\bf 424}, 1022 (2003).

\bibitem{Bevan} T.D.C. Bevan, A.J. Manninen, J.B. Cook, H. Alles, J.R. Hook, H.E. Hall, {\it J. Low Temp. Phys.}, {\bf 109}, 423 (1997); {\it Phys. Rev. Lett.} {\bf 74}, 750 (1995).

\bibitem{FlightTime} A.P. Finne, V.B. Eltsov, R. Blaauwgeers, Z. Janu, M. Krusius, L. Skrbek,  {\it J. Low Temp. Phys.}  {\bf 134}, 375 (2004).

\bibitem{NeutronTurbulence} A.P. Finne, S. Boldarev, V.B. Eltsov, M. Krusius, {\it J. Low Temp. Phys.} {\bf 135}, 479 (2004).

\bibitem{TurbPhaseDiagram} A.P. Finne, S. Boldarev, V.B. Eltsov, M. Krusius, {\it J. Low Temp. Phys.}, submitted (2005).

\bibitem{ToBePublished} A.P. Finne {\it et al.}, to be published.

\bibitem{VortexSheet} V.B. Eltsov, R. Blaauwgeers, N.B. Kopnin, M. Krusius, J.J. Ruohio, R. Schanen, E.V. Thuneberg, {\it Phys. Rev. Lett.}  {\bf 88}, 065301 (2002).

\bibitem{Sonin} E.B. Sonin, {\it Rev. Mod. Phys.} {\bf 59}, 87 (1987).

\bibitem{Glaberson} R.M. Ostermeier and W.I. Glaberson, {\it J. Low Temp. Phys.} {\bf 21}, 191 (1975); W.I. Glaberson, W.W. Johnson, and R.M. Ostermeier, {\it Phys. Rev. Lett.} {\bf 33}, 1197 (1974).

\bibitem{volovik2reg} G.E. Volovik, {\it JETP Lett.} {\bf 78} 533 (2003).

\bibitem{VorFlow} L. Skrbek, R. Blaauwgeers, V.B. Eltsov, A.P. Finne, N.B. Kopnin, and M. Krusius, {\it Physica} B {\bf 329-333}, 106 (2003).

\bibitem{Rob} R. Blaauwgeers, V.B. Eltsov, A.P. Finne, M. Krusius, {\it Physica} B {\bf 329-333}, Pt. 1, 93 (2003).

\bibitem{AB-InterfaceShape} V.B. Eltsov, R. Blaauwgeers, A.P. Finne, M. Krusius, J.J. Ruohio, G.E. Volovik, {\it Physica} B {\bf 329-333}, Pt. 1, 96 (2003).

\bibitem{FootNote} The locations and shapes of the AB interfaces\cite{AB-InterfaceShape} in Fig.~\protect\ref{ExpSetUp} have been calculated at $P=29.0\,$bar, $\Omega = 0$, and $T=0.60\,T_{\rm c} = 1.46\,$mK, where the B$\rightarrow$A transition field is $H_{\rm AB}=340\,$mT. At the barrier current $I_{\rm b} =8.0\,$A, the interface is almost flat and at a distance $z= \pm 8\,$mm from the mid plane of the magnet. At the same $I_{\rm b} =8.0\,$A at $0.40\,T_{\rm c}$, the interface is more curved and at $z= \pm 6\,$mm.

\bibitem{DensityAnisotropy} J.S. Korhonen, Yu.M. Bunkov, V.V. Dmitriev, Y. Kondo, M. Krusius, Yu.M. Mukharskiy, \"U. Parts, and E.V. Thuneberg, {\it Phys. Rev.} B, {\bf 46}, 13983 (1992).

\bibitem{CF-Textures} J.S. Korhonen, A.D. Gongadze, Z. Janu, Y. Kondo, M. Krusius, Yu.M. Mukharsky, and E.V. Thuneberg, {\it Phys. Rev. Lett.} {\bf 65}, 1211 (1990).

\bibitem{Hakonen} P.J. Hakonen, M. Krusius, M.M. Salomaa, R.H. Salmelin, J.T. Simola, A.D. Gongadze, G.E. Vachnadze, and G.A. Kharadze, {\it J. Low Temp. Phys.} {\bf 76}, 225 (1989).

\bibitem{Ahonen} A.I. Ahonen, M. Krusius, and M.A. Paalanen, {\it J. Low Temp. Phys.} {\bf 25}, 421 (1976).

\bibitem{AB-VortexLayer} R. H\"{a}nninen, R. Blaauwgeers, V.B. Eltsov, A.P. Finne, M. Krusius, E.V. Thuneberg, G.E. Volovik, {\it Phys. Rev. Lett.} {\bf 90}, 225301 (2003).

\bibitem{KH-Theory} G.E. Volovik, {\it JETP Lett.} {\bf 75}, 491 (2002).


\bibitem{Simulation} M. Tsubota, A. Mitani, {\it J. Low Temp. Phys.} this issue; private communication.

\bibitem{FootNote1} We assume the following values at 10.2\,bar [29.0\,bar]: $\sigma_{\rm AB}=3.36\,(1-T/T_{\rm c})^{3/2} \cdot 10^{-5}\,$erg/cm$^2$ [$\sigma_{\rm AB}=4.88\,(1-T/T_{\rm c})^{3/2} \cdot 10^{-5}\,$erg/cm$^2$], $\rho_{\rm sB}(H\!=\!0)=0.055\,$g/cm$^3$ [$0.047\,$g/cm$^3$], and $H=H_{\rm AB}= 449\,$mT [$367\,$mT]. For better agreement we take $\rho_{\rm sB}(H\!=\!H_{\rm AB}) \approx 1.2\,\rho_{\rm sB}(H\!=\!0)$ [$\rho_{\rm sB}(H\!=\!H_{\rm AB}) \approx 1.15\,\rho_{\rm sB}(H\!=\!0)$].

\end{thebibliography}
\end{document}